\documentclass[aps,prb,twocolumn,superscriptaddress,floatfix,10pt]{revtex4-1}
\usepackage[utf8]{inputenc}
\usepackage{tabularx}
\usepackage{array}
\usepackage[dvipsnames]{xcolor}
\usepackage{graphicx}
\usepackage{multirow}
\usepackage{longtable}
\usepackage{amsmath}
\usepackage{amssymb}
\usepackage{textcomp}
\usepackage{color}
\usepackage{gensymb}%\degree must load package textcomp before
\usepackage{times}
\newcommand{\modif}[1]{\textcolor{black}{#1}}

\begin{document}

\title{Anomalous temperature-dependent magnetization in the nearly collinear antiferromagnet Y$_2$Co$_3$}
\author{Yunshu Shi}
\affiliation{Department of Physics and Astronomy, University of California, Davis, California 95616, USA}
\author{Huibo Cao}
\affiliation{Neutron Scattering Division, Oak Ridge National Laboratory, Oak Ridge, Tennessee 37831, USA}
\author{Hung-Cheng Wu}
\affiliation{Institute of Multidisciplinary Research for Advanced Materials, Tohoku University, Sendai 980-8577, Japan}
\affiliation{Department of Physics, National Sun Yat-sen University, Kaohsiung, 80424, Taiwan}
\author{Li Yin}
\affiliation{Materials Science and Technology Division, Oak Ridge National Laboratory, Oak Ridge, Tennessee 37831, USA}
\author{Neil Harrison}
\affiliation{Los Alamos National Laboratory, Los Alamos, New Mexico 87545, USA}
\author{David S. Parker}
\affiliation{Materials Science and Technology Division, Oak Ridge National Laboratory, Oak Ridge, Tennessee 37831, USA}
\author{Tushar Bhowmick}
\affiliation{Department of Physics and Astronomy, University of Utah, Salt Lake City, Utah 84112, USA}
\author{Tessa McNamee}
\affiliation{Department of Physics and Astronomy, University of Utah, Salt Lake City, Utah 84112, USA}
\author{Fatemeh Safari}
\affiliation{Department of Physics and Astronomy, University of Utah, Salt Lake City, Utah 84112, USA}
\affiliation{Department of Physics, University of Illinois Chicago, Chicago, Illinois 60607, USA}
\author{Sergey L. Bud'ko}
\affiliation{Department of Physics and Astronomy, Iowa State University, Ames, Iowa 50011, USA}
\affiliation{Ames National Laboratory, US DOE, Iowa State University, Ames, Iowa 50011, USA}
\author{James C. Fettinger}
\affiliation{Department of Chemistry, University of California, Davis, California 95616, USA}
\author{Susan M. Kauzlarich}
\affiliation{Department of Chemistry, University of California, Davis, California 95616, USA}
\author{Peter Klavins}
\affiliation{Department of Physics and Astronomy, University of California, Davis, California 95616, USA}
\author{Dmitry Popov}
\affiliation{HPCAT, X-ray Science Division, Argonne National Laboratory, Lemont, Illinois 60439, USA}
\author{Ravhi Kumar}
\affiliation{Department of Physics, University of Illinois Chicago, Chicago, Illinois 60607, USA}
\author{Russell J. Hemley}
\affiliation{Department of Physics, University of Illinois Chicago, Chicago, Illinois 60607, USA}
\affiliation{Department of Chemistry, University of Illinois Chicago, Chicago, Illinois 60607, USA}
\author{Shanti Deemyad}
\affiliation{Department of Physics and Astronomy, University of Utah, Salt Lake City, Utah 84112, USA}
\author{Taku J. Sato}
\affiliation{Institute of Multidisciplinary Research for Advanced Materials, Tohoku University, Sendai 980-8577, Japan}
\author{Paul. C. Canfield}
\affiliation{Department of Physics and Astronomy, Iowa State University, Ames, Iowa 50011, USA}
\affiliation{Ames National Laboratory, US DOE, Iowa State University, Ames, Iowa 50011, USA}
\author{Valentin Taufour}
\email{vtaufour@ucdavis.edu}
\affiliation{Department of Physics and Astronomy, University of California, Davis, California 95616, USA}
%\date{\today}

\begin{abstract}

Y$_2$Co$_3$ is a newly discovered antiferromagnetic (AFM) compound with distorted kagome layers. Previous investigations via bulk magnetization measurements suggested a complex noncollinear magnetic behavior, with magnetic moments primarily anti-aligned along the $b$ axis and some canting towards the $ac$ plane. In this study, we report the magnetic structure of Y$_2$Co$_3$ to be an A-type AFM structure with ferromagnetic (FM) interactions within the distorted kagome plane and an interplane antiferromagnetic interaction, as determined by single-crystal neutron diffraction. The magnetic moments align along the $b$ axis, with minimal canting towards the $c$ axis, at odds with the previous interpretation of bulk magnetization measurements. The magnetic moments on the two distinct Co sites are [0, -0.68(3), 0]\,$\mu_B$ and [0, 1.25(4), 0.07(1)]\,$\mu_B$. We attribute the previously reported ``noncollinear" behavior to the considerable temperature dependence of itinerant AFM exchange interactions, induced by thermal contraction along the $b$ axis. Additionally, our examination of lattice constants through pressure studies reveals compensating effects on FM and AFM interactions, resulting in negligible pressure dependence of $T_\textrm{N}$.

\end{abstract}

\maketitle

\section{Introduction}

Antiferromagnetism in kagome lattices has attracted significant interest within the condensed matter community  because of the potential emergence of exotic quantum states from the unique lattice structure and frustrated magnetic ordering at low temperatures. Novel magnetic and transport properties may arise from these phenomena, such as quantum spin liquid states~\cite{Han2012Fractionalized}, flat bands~\cite{Sales2021PRM}, and the anomalous Hall effect~\cite{Nakatsuji2015Large, Ma2021}. The R$_2$Co$_{3-x}$Si$_x$ (R = Y, La, Ce, Pr, Nd, and Gd; $0 \leq x < 0.5$) family, featuring the La$_2$Ni$_3$-type structure, hosts distorted kagome lattices within the $ac$ plane. Most compounds in this family are ferromagnetic (FM)~\cite{Mahon2018R2Co3-xSix, Tence2014Stabilization} or ferrimagnetic~\cite{Byland2021Statistics}, with magnetic moments originating from both rare-earth elements and cobalt. 

Y$_2$Co$_3$ and La$_2$Co$_3$ are among the richest Co-based antiferromagnets, exhibiting relatively high Néel temperatures of 252\,K and 315\,K~\cite{ShiRobust2021PRB,gignoux1985antiferromagnetism}, respectively. Prior research on single-crystal Y$_2$Co$_3$~\cite{ShiRobust2021PRB} revealed antiferromagnetic ordering with magnetic moments primarily aligned along the $b$ axis. Bulk magnetization measurements exhibited temperature-dependent susceptibility along the hard axes and a relatively large nonzero susceptibility along the easy axis at low temperatures. This behavior was interpreted as a complex noncollinear magnetic structure. In contrast, we report single-crystal neutron diffraction measurements, which indicate an almost collinear A-type antiferromagnetic structure.

The observed discrepancy is carefully examined in this paper. The temperature dependence of perpendicular magnetic susceptibility is explained by the relatively large temperature dependence of the AFM interaction owning to the thermal contraction below the transition temperature. By comparing the Gr\"{u}neisen parameter of Y$_2$Co$_3$ with other Y-Co compounds~\cite{Brouha1973Pressure}, we demonstrate that the change in lattice constants across the wide temperature range may increase the AFM exchange interaction by up to 40\%, consequently decreasing the perpendicular magnetic susceptibility by 30\%, resembling that of a noncollinear antiferromagnet. We further reveal that the intraplane ferromagnetic interaction and interplane antiferromagnetic interaction have compensating dependencies on lattice parameters. This finding may also account for the previously reported small pressure dependence of $T_\textrm{N}$ and the ``robustness" of antiferromagnetism in this compound. Nevertheless, the presence of nonzero parallel magnetic susceptibility remains puzzling. We rule out sample misalignment and minimal canting moment along the $c$ axis as possible explanations. We hypothesize that residual easy axis magnetic susceptibility may originate from the itinerant antiferromagnetic fluctuations, as supported by both experimental evidence and spin-polarized density function theory (DFT) calculations.

\section{Methods}
Single crystals of Y$_2$Co$_3$ were synthesized using the flux growth method~\cite{ShiRobust2021PRB}. A starting composition of Y$_{51.5}$Co$_{48.5}$ was arc-melted and sealed in a clean tantalum crucible with a tantalum filter~\cite{Jesche2014PM}. The tantalum
assembly was sealed in a silica tube with partial argon
pressure. The sample was then heated up to 1150\,$\celsius$ within 4 hours and held for 5\,hours, quickly cooled down to 945\,$\celsius$ and slowly cooled down to 825\,$\celsius$ within 133\,hours. 

% Single crystals La$_2$Co$_3$ were grown with the similar flux growth method using tantalum tube. A starting composition of La$_{52.5}$Co$_{47.5}$ was arc-melted. Following the same experiment method, the sample was heated up to 1150 $\celsius$, quickly cool down to 720 $\celsius$ and then slowly cool down to 620 $\celsius$ within 260 hours. 

Neutron single-crystal diffraction was performed to determine the magnetic structure at the DEMAND instrument (HB-3A beamline) in its four-circle mode at the High Flux Isotope Reactor of the Oak Ridge National Laboratory~\cite{cryst9010005}. A neutron wavelength of 1.542\,\AA\ was used from a bent perfect Si-220 monochromator~\cite{Chakoumakosko5139}. The magnetic and nuclear peaks are collected at 5 K. The Bilbao Crystallography Server was used for the magnetic symmetry analysis~\cite{Perez-Mato2015}. Magnetic structure refinements were carried out with the FullProf Suite program~\cite{RODRIGUEZCARVAJAL199355}.

To determine the weak magnetic component along the $c$ axis, single-crystal neutron diffraction measurements were also performed using the GPTAS (4G) triple-axis thermal neutron spectrometer in two-axis mode (without an analyzer) at JRR-3, Tokai, Japan~\cite{Nawa2024GPTAS}. The incident neutron was selected at 14.7\,meV using a vertically focusing PG 002 monochromator. To eliminate higher harmonic neutron contributions, two PG filters were inserted, one after the monochromator and another after the sample. A single crystal of Y$_2$Co$_3$, weighing approximately 30\,mg, was aligned in the ($h0l$) plane as scattering planes. The crystal was sealed in a standard aluminum sample can with $^{4}\text{He}$ exchange gas. Cooling was achieved using a Gifford-McMahon (GM) cryostat equipped with a closed-cycle $^{4}\text{He}$ refrigerator. $\theta$$-$2$\theta$ scans were performed to obtain the $|F|^2$ table, after correcting for the Lorentz-polarization factor.

Magnetization measurements were performed with a Magnetic Property Measurement System (MPMS, Quantum Design) in the temperature range 2\,K$-$300\,K with the applied magnetic fields up to 7\,T. 
The angle-dependent magnetization measurements at constant temperatures
2, 5, and 10\,K and applied field 20\,kOe were performed using a horizontal
rotator option of a Quantum Design MPMS3 SQUID magnetometer. The sample
was aligned with the rotation axis visually, using characteristic
features of its morphology and glued to the rotator platform with a
small amount of GE-7031 varnish. The direction of the magnetic field was
changing in the $ab$ and $ac$ crystallographic planes. Empty rotator
measurements were performed at the same $H$, $T$ conditions as well, but
since the angular change of the empty rotator response was ~ three
orders of magnitude smaller than that of the sample, no background
subtraction was performed.

Resistivity, Hall resistivity, and heat capacity measurements were carried out with a Physical Property Measurement System (PPMS, Quantum Design).  Resistivity data was measured using the four-probe method with the current in the $ac$ plane.The Hall resistivity measurement was applied using the five-probe method. The magnetic field was swept from 9\,T to $-$9\,T to remove the longitudinal signal, which is an even function of the field.

\section{Result and Discussion}
\subsection{Magnetic structure}
\begin{figure}[!htb]
	\centering
	\includegraphics[width=\linewidth]{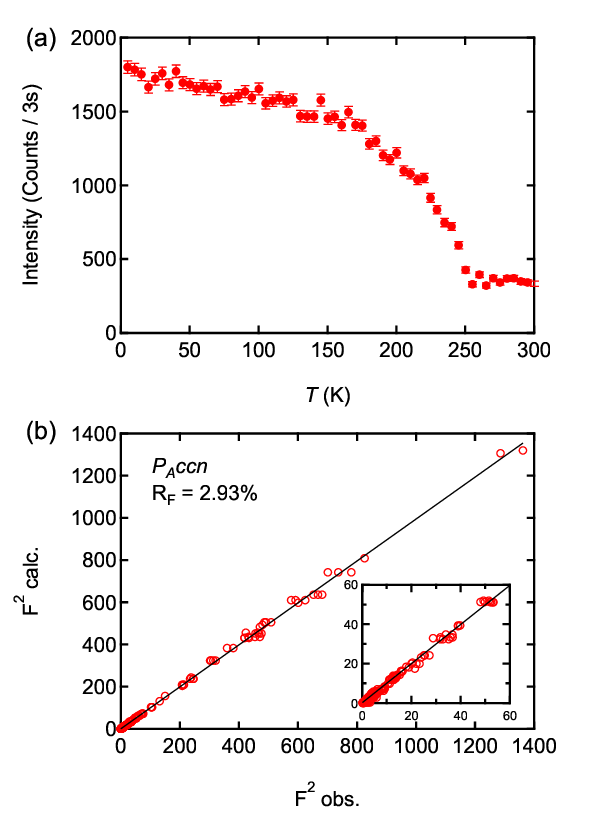}
	\caption{(a) The  $(0 1 \overline{1})$ magnetic Bragg peak intensity as a function of temperature, corresponding to the squared magnetic order parameter.  (b) The calculated magnetic structure factor ($|F|^2_{calc.}$) derived from the refinement, assuming the magnetic space group of $P_Accn$, compared to the observed magnetic structure factor ($|F|^2_{obs.}$) obtained at 1.5\,K. The inset illustrates the distribution on the lower side of $|F|^2$. The linear relation between $|F|^2_{calc.}$ and $|F|^2_{obs.}$ indicates a reliable refinement. \label{IvT}}  
\end{figure}

Figure~\ref{IvT}(a) exhibits the selected magnetic $(0 1 \overline{1})$ Bragg reflection as a function of temperature from 5\,K to 300\,K. Below the N\'{e}el temperature ($T_\mathrm{N} \sim 252\,\text{K}$) a pure magnetic signal develops at the $(0 1 \overline{1})$ Bragg position indicating that the magnetic moments have a component along a direction perpendicular to $(0 1 \overline{1})$. This result is compatible with the moments along the $b$ axis. A transition temperature of around 250\,K is consistent with that obtained from single-crystal magnetization measurement.

\begin{figure}[!htb]
	\centering
	\includegraphics[width=\linewidth]{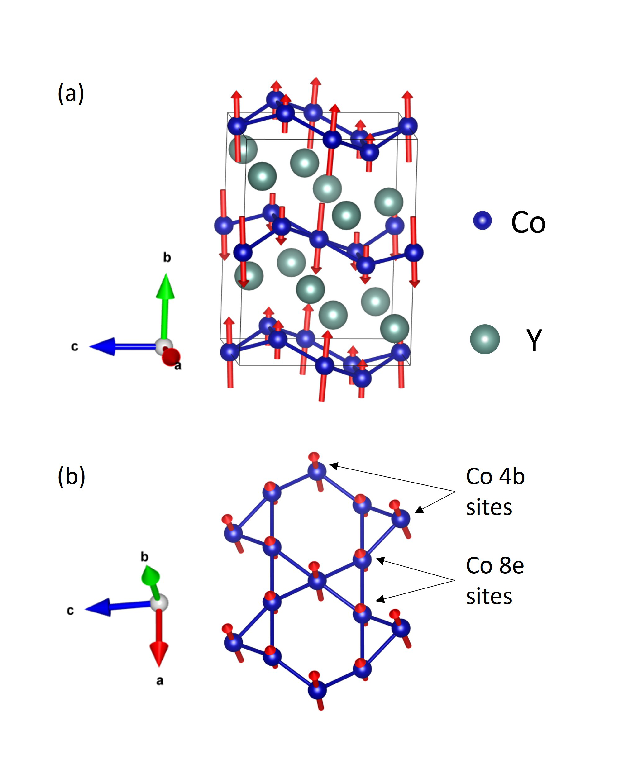}
	\caption{(a) Illustration of the magnetic Structure of Y$_2$Co$_3$. (b) The magnetic structure of Y$_2$Co$_3$ in one Co layer.\label{MagneticStructure_Neutron}}  
\end{figure}

The magnetic symmetry analysis using Bilbao Crystallography Server with the magnetic propagation vector of $\vec{k}$ = $(0 1 0)$ and Wyckoff positions of two Co sites being 4b and 8e, yields eight possible magnetic space group candidates. The magnetic space group $P_Accn$ (BNS notation: No. 56.374) is the only model that satisfies the observed magnetic reflections in the refinement, as shown in Fig.~\ref{IvT}(b). The resolved magnetic structure is displayed in 
 Fig.~\ref{MagneticStructure_Neutron}(a), where the Co (8e) site has a magnetic component only along the $b$ axis [$\textbf{m}_{8e} = (0, -0.68(3), 0)\,\mu_B$] and the Co (4b) site has a major magnetic moment along the $b$ axis and a minor component along the $c$ axis [$\textbf{m}_{4b} = (0, 1.25(4), 0.03(8))\,\mu_B$]. However, the refine $c$ component is too small to be determined within the instrument’s resolution. 

%The magnetic structure of Y$_2$Co$_3$ is shown in figure \ref{MagneticStructure_Neutron}. The optimal fit reveals a magnetic space group of $P_Accn$. With this magnetic space group, the Co 8e site (shown in Figure~\ref{MagneticStructure_Neutron} (b) only permits magnetic moment along the $b$ axis with $m_{8e} = (0, 0.68(3), 0)\,\mu_B$. While the Co 4b site allows magnetic moment components along the $b$ and $c$ directions with $m_{4b}= (0, 1.25(4),0.03(8))\,\mu_B$, however, the observed $c$ component value is too small and within the instrument's resolution. This magnetic space group is validated by the agreement between the observed structure factor $F_{obs}$ and the calculated one $F_{calc}$ presented in Figure~\ref{IvT} (b).  

The overall magnetic moments appear relatively small compared to the local moment of Co (1.7\,$\mu_{B}$/Co) or the effective moment of $\mu_{eff}=2.513(4)$\,$\mu_{B}$/Co~\cite{ShiRobust2021PRB} obtained from the Curie-Weiss fitting above the transition temperature, suggesting some degree of itinerant character in Y$_2$Co$_3$. Within each layer of the $ac$ plane, Co moments order ferromagnetically, while the antiferromagnetic alignment is along the $b$ axis between layers of Co atoms separated by Y atoms. The intraplane ferromagnetic interactions discovered in the neutron diffraction experiment coincide with the positive Curie-Weiss temperature obtained by Curie-Weiss fitting~\cite{ShiRobust2021PRB}, which will be discussed in details later.

\begin{figure}[!htb]
	\centering
	\includegraphics[width=\linewidth]{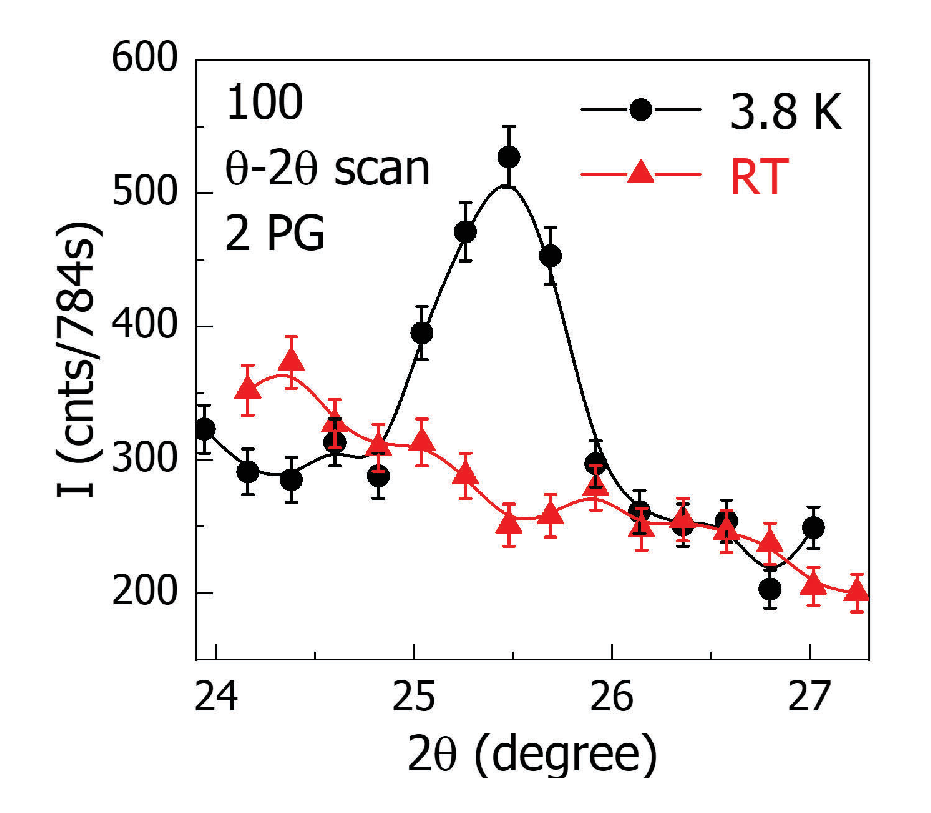}
	\caption{$\theta$$-$2$\theta$ scans performed at 3.8 K and room temperature were used to obtain the integrated intensity of selected 100 reflection, indicating the presence of a magnetic component along the $c$ axis. The magnetic moment of Co (4b) along the $c$ axis was calculated to be 0.07(1) $\mu_B$. It is worth noting that magnetic moment along the $b$ axis does not contribute to any magnetic reflection at 100 reflection. The configuration of GPTAS (4G) instrument was set to 40’-PG-40’-40’-PG-open, in two-axis mode.}\label{theat-two theta}  
\end{figure}

Simulations of magnetic structure factors indicate that the magnetic moment along the $c$ axis at the Co (4b) site results in specific magnetic reflection peaks along $(h00)$, following the extinction rule $h = 2n+1$. To further confirm the weak tilting moment along the $c$ axis, the sample was aligned to the $(h 0 l)$ plane as a scattering plane. As shown in Fig.~\ref{theat-two theta}, at room temperature, there is no observable reflection at $(1 0 0)$ reflection, consistent with the space group of $Cmce$ (No. 64). At 3.8\,K, the magnetic $(1 0 0)$ reflection is clearly observed, indicating the formation of a magnetic moment along the $c$ axis. A scale factor was determined from the $(2 0 0)$ nuclear reflection. The calculated moment size based on $(2 0 0)$ nuclear reflection and $(1 0 0)$ magnetic reflection at 3.8\,K was 0.07(1) $\mu_B$, \modif{which provides a more precise determination than the previously mentioned value of 0.03(8)\,$\mu_B$ from magnetic structure refinement.} 

During the extended scan at selected nuclear reflection and magnetic reflection using the 4G instrument, it is noteworthy that we detected measurable intensity at the $(2 0 1)$ and $(2 0 3)$ reflections at room temperature (in the paramagnetic phase). The occurrence of these reflections is unusual as they are not allowed by the $Cmce$ space group. This suggests that their origin could stem from extrinsic factors, such as higher-harmonic neutrons or double reflections, or possibly from intrinsic symmetry lowering because of the large crystal size. Despite this, the intensity ratio between the strongest $(2 0 0)$ reflection and weak $(2 0 1)$ reflection is only approximately 0.05\%. Consequently, the presence of the $(201)$ reflection is unlikely to significantly affect the analysis of the magnetic structure in this paper.

\subsection{Temperature-dependent bulk mnagnetization}

\begin{figure}[!htb]
	\centering
	\includegraphics[width=\linewidth]{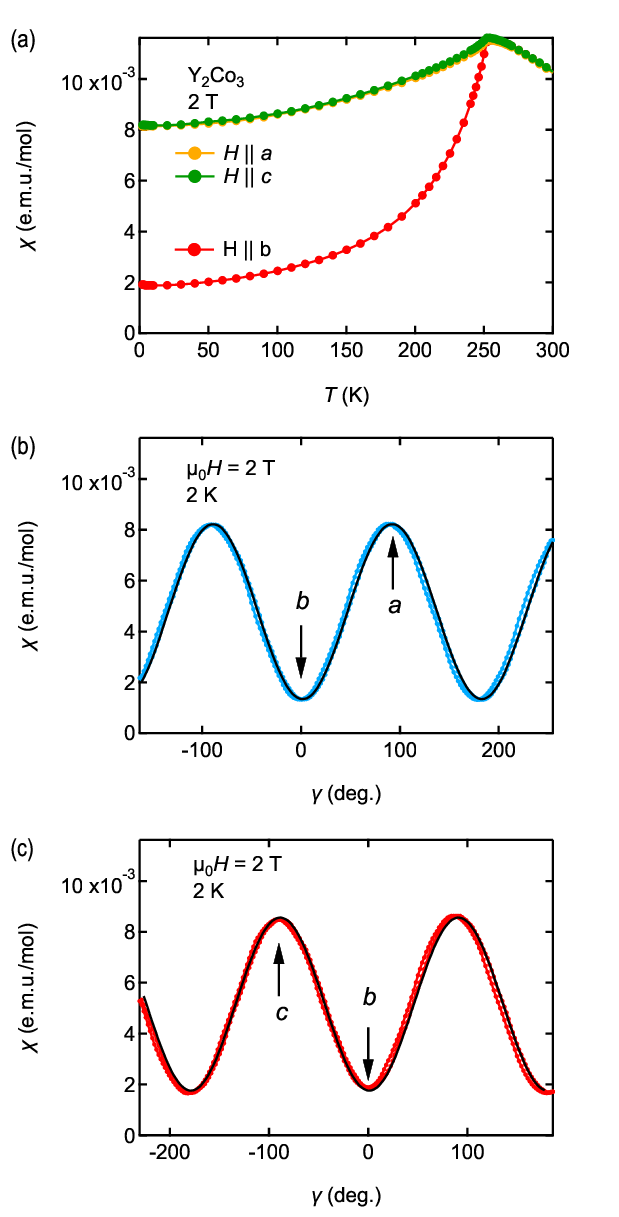}
	\caption{(a) Susceptibility as a function of temperature of single crystal Y$_2$Co$_3$ with the magnetic field along the $a$, $b$, and $c$ axes. [(b), (c)] The magnetic susceptibility as a function of $\gamma$, where $\gamma$ is the angle between the magnetic field and the $b$ axis in the $b-a$ plane and $b-c$ plane.	\label{bulk_magnetization_combined}}  
\end{figure}

The single-crystal neutron diffraction confirms the major magnetic moments along the $b$ axis and reveals an almost collinear magnetic structure. In a collinear Heisenberg antiferromagnet with local moments, the magnetic susceptibility is anticipated to be nearly zero at 0\,K along the ordering axis. When the applied field is perpendicular to the ordering axis, the magnetic susceptibility is expected to be temperature independent~\cite{Johnston2012MagneticPRL}. Examples supporting this theoretical model include Ni$_3$TeO$_6$, GdNiGe$_3$ and MnF$_2$ \cite{Johnston2012MagneticPRL, ivkovi2010Ni3TeO6, MUN2010RNiGe3}. However, the bulk magnetization measurements on Y$_2$Co$_3$ show a relatively large nonzero susceptibility at 2\,K along the ordering $b$ axis. Moreover, when the magnetic field is applied perpendicular to the ordering axis, the susceptibility of Y$_2$Co$_3$ decreases as the temperature drops [Fig.~\ref{bulk_magnetization_combined}(a)].

If this bulk susceptibility is attributed to canting, we first try to roughly estimate the canting angle $\alpha$ using $\chi_b$ at $T=2$\,K. By decomposing the applied magnetic field into the directions parallel and perpendicular to the magnetic moment, and summing the corresponding susceptibility, we get the formula $\sin^2\alpha = \frac{\chi_b(0)}{\chi_b(T_\textrm{N})}$ (see Appendix C). This yields $\alpha=23^{\circ}$, a value significantly larger than the observed canting angle of the magnetic moments on the Co (4b) sites, which is approximately 3$^{\circ}$ as determined by neutron diffraction. Moreover, the temperature dependence in $\chi_a$ and $\chi_c$ cannot be attributed to the $c$ component of the magnetic moment, as magnetization along the $a$ and $c$ axes exhibits identical behavior.

We then check whether a misalignment would be a possible explanation utilizing angular-dependent magnetization measurement. Figures~\ref{bulk_magnetization_combined}(b) and \ref{bulk_magnetization_combined}(c) show the magnetization as a function of the tilted angle from the $b$ axis toward the $a$ and $c$ axes.  The experiment data aligns well with a $\sin^2\gamma$ function (solid line), where $\gamma$ is the angle from the $b$ axis. This indicates that the magnetization reaches its minimum exclusively along the $b$ axis, as opposed to other directions. This outcome further confirms that the temperature dependence of the susceptibility cannot be attributed to sample misalignment, which refers to the possibility that the sample might not be perfectly aligned along the intended crystallographic axis during the measurement. Such a misalignment would mean that our measurements could contain contributions from multiple crystallographic directions.

\subsection{The effect of lattice contraction on exchange interaction}

Trying to understand the temperature dependence of anisotropic magnetic susceptibility, we use the molecular field theory (MFT) to analyze the two different types of exchange interactions and the previously reported magnetization behavior in more details. In Y$_2$Co$_3$, the interplane antiferromagnetic (AFM) exchange energy $J_b$ and the intraplane ferromagnetic (FM) exchange energy $J_{ac}$ can be estimated from $T_\textrm{N}$ and the Curie-Weiss temperature following MFT. In this framework, a positive exchange energy value indicates FM interactions, while a negative value corresponds to AFM interactions. According to MFT, for a lattice with identical spins, the Curie Weiss temperature $\theta_P$ and N\'{e}el temperature $T_\textrm{N}$ can be expressed as follows~\cite{Johnston2012MagneticPRL, JohnstonPRB2011Magnetic, Blundell2001Magnetism1}:

\begin{subequations}
\begin{align}
 \theta_P &= \frac{S(S+1)}{3k_B}\sum_j J_{ij} \\
 &= \frac{S(S+1)}{3k_B}(z_{ac}J_{ac}+z_bJ_b) \label{eqn:1a}
\end{align}
\\
\begin{align}
 T_\textrm{N} &= \frac{S(S+1)}{3k_B}\sum_j J_{ij}\cos{\phi_{ij}}\\
 &= \frac{S(S+1)}{3k_B}(z_{ac}J_{ac}-z_bJ_b) \label{eqn:1b}
\end{align}
\end{subequations}

In these equations, the sums are taken over the nearest neighbors $j$ around the $i$th spin, where $J_{ij}$ represents the exchange coupling between the $i$th and $j$th spins. Consequently, the summation can be simplified as the product of the exchange coupling and the number of nearest neighbors, with $z_{ac}$ denoting the number of nearest neighbors within the $ac$ plane exhibiting FM exchange couplings, and $z_b$ representing the number of interplane nearest neighbors along the $b$ axis with AFM exchange couplings.

In the case of Y$_2$Co$_3$, the two different cobalt sites are not crystallographically equivalent, nevertheless, the average molecular fields can still be estimated. For an antiferromagnetic compound, there exist molecular field couplings between two different sublattices ($\lambda_d$) and within the same sublattice ($\lambda_s$). In Y$_2$Co$_3$, $\lambda_d$ represents the interplane AFM coupling and $\lambda_s$ represents the intraplane FM coupling. The expressions of these two terms are as follows~\cite{JohnstonPRB2011Magnetic, Kittel2004}:

\begin{subequations}
\begin{eqnarray}
 \lambda_d = \frac{2z_bJ_b}{Ng^2\mu_B^2} \label{eqn:2a}\\
 \lambda_s = \frac{2z_{ac}J_{ac}}{Ng^2\mu_B^2\label{eqn:2b}}
\end{eqnarray}
\end{subequations}

Thus, Eqs.~(\ref{eqn:1a}) and (\ref{eqn:1b}) can be rewritten as

\begin{subequations}
\begin{eqnarray}
 \theta_P = \frac{Ng^2\mu_B^2S(S+1)}{6k_B}(\lambda_s+\lambda_d) = \frac{C}{2}(\lambda_s+\lambda_d)\\
 T_\textrm{N} = \frac{Ng^2\mu_B^2S(S+1)}{6k_B}(\lambda_s-\lambda_d) = \frac{C}{2}(\lambda_s-\lambda_d)
\end{eqnarray}
\end{subequations}
where $N$ is number of spins and $C = \frac{Ng^2\mu_B^2S(S+1)}{3k_B}$ is the Curie constant. With $T_\textrm{N}=252$\,K and $\theta_P=59.5$\,K, the intra- and interplane molecular fields can be estimated to be $C\lambda_s\approx$ 312\,K and  $C\lambda_d\approx$ -192\,K.

Along the perpendicular direction, the magnetic susceptibility can be expressed as (see Appendix~B)

\begin{equation}
    \chi_{\perp}(T) = \frac{2M^2(T)}{2|\lambda_d|M^2(T)+K(T)}\label{eqn7}
 \end{equation}

In this equation, $M(T)$ is the spontaneous magnetization of the sublattices, and $K(T)$ is the temperature-dependent magnetocrystalline anisotropy. At 2\,K,  $K=5.74$\,J/mol\,Co~\cite{ShiRobust2021PRB} is much smaller than the exchange term $|\lambda_d|M^2$ $\approx200$\,J/mol\,Co using the value of $C\lambda_d= -192$\,K, the Curie constant obtained from the Curie-Weiss fitting~\cite{ShiRobust2021PRB} and $M = \frac{1}{2}N_Ag\mu_BS$. Consequently, the temperature dependence of the $K(T)$ term (see Appendix~D) would only cause a change in $\chi_\perp$ of less than 3\%, indicating that the large drop of perpendicular magnetic susceptibility can not be explained by magnetocrystalline anisotropy. Equation~(\ref{eqn7}) can be simplified as

\begin{equation}
    \chi_{\perp}(T) = \frac{1}{|\lambda_d|}\label{eqn8}
\end{equation}

\begin{figure}[!htb]
	\centering
	\includegraphics[width=\linewidth]{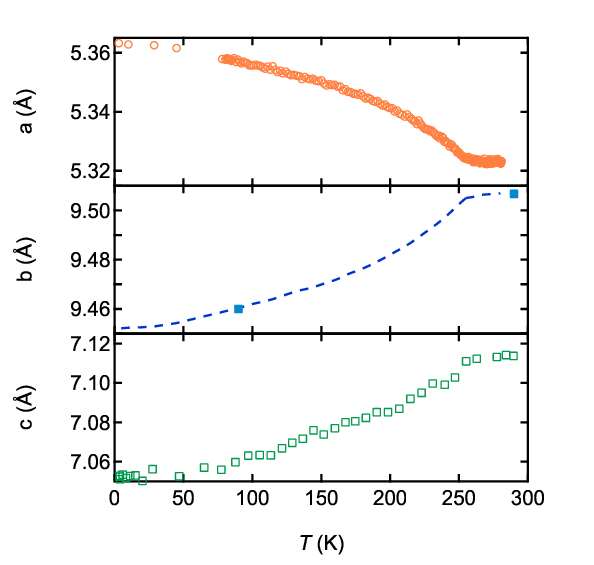}
	\caption{The lattice parameter as a function of $a$, $b$, and $c$. The markers are experimental data, and the dashed line is the interpolated curve. Because of limited beam time, measurements were confined to $a$ and $c$ parameters during the cooling and warming procedures, with the behavior of $b$ deduced through interpolation based on the two data points at 90\,K and 290\,K, and the temperature dependence of $a$.
	\label{lattice_parameter}} 
\end{figure}

We propose that the decrease in the perpendicular magnetic susceptibility is attributed to the enhancement of the AFM exchange interaction. This hypothesis is preliminarily evaluated by comparing the exchange coupling constants at high and low temperatures. As previously noted, the $C\lambda_d$ value near the transition temperature is $-$192\,K, while at 2\,K, this value is estimated to be approximately $-$303\,K, based on the previously reported spin-flop field and the magnetocrystalline anisotropy energy~\cite{ShiRobust2021PRB}.  A more detailed discussion of this estimation is presented in Appendix~D. The smaller absolute value of $\lambda_d$ at higher temperature indicates that the exchange coupling strength increases as the temperature decreases. We propose that this is caused by thermal contraction along the $b$ axis upon cooling. Figure~\ref{lattice_parameter} shows the change of lattice parameters $a$, $b$, and $c$ as a function of temperature obtained from single-crystal neutron diffraction. Because of limited beam time, measurements were confined to $a$ and $c$ parameters during the cooling and warming procedures, with the behavior of $b$ deduced through interpolation based on the two data points at 90\,K and 290\,K and the temperature dependence of $a$. We observe that the thermal contraction exhibits a distinct change in slope at the AFM transition temperature, indicating strong coupling between the AFM ordering and lattice parameters. This suggests that thermal contraction upon cooling could enhance the antiferromagnetic coupling, resulting in the observed temperature dependence of $\chi_{\perp}$. This explanation can apply more generally to other materials. For example, other Co-based antiferromagnetic materials with collinear magnetic structures, such as KCoF$_3$, K$_2$CoF$_4$, and Co$_2$Mo$_3$O$_8$, also show a slight decrease in perpendicular magnetic susceptibility upon cooling~ \cite{Tsuda1978,BREED1969205, Tang2019PRBCo2Mo3O8, McAlister1983JMMM}. %The observed thermal contraction upon cooling below $T_\textrm{N}$ is indicative of magnetic ordering's influence.
Notably, upon cooling, there is an expansion along the $a$ axis leading to an increased distortion of the kagome lattice in the $ac$ plane.

\begin{figure}[!htb]
	\centering
	\includegraphics[width=\linewidth]{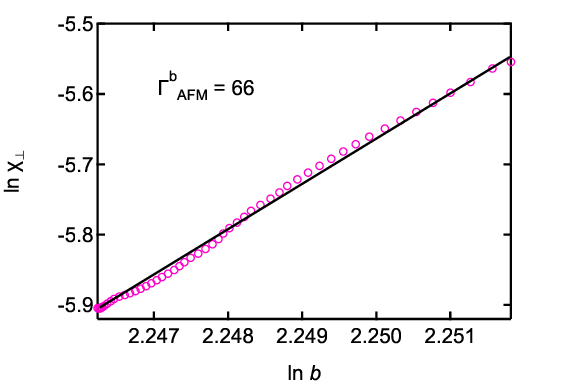}
	\caption{$\ln{\chi_{\perp}}$ as a function of $\ln{b}$ below $T_\textrm{N}$ showing linear relationship.
	\label{Gruneisen}}
\end{figure}

To explore the relationship between the AFM exchange interaction and the lattice parameter $b$, we calculate the volume Gr\"{u}neisen parameter, $\Gamma^i = -d\ln |J|/d\ln a_i$~\cite{Hardy2009PRB, Brouha1973Pressure} of the AFM component, where $J$ is the exchange energy, and $a_i$ represents the lattice constant of the orthorhombic axes and/or the volume.  Referring to Eqs.~(\ref{eqn:2a}) and (\ref{eqn8}), we derive $\Gamma^b = d\ln \chi_{\perp}/d\ln b\approx 66$ (see Fig.~\ref{Gruneisen}). The positive sign indicates that the AFM coupling becomes stronger as the temperature decreases, which is also seen in other Co-based antiferromagnetic compounds, such as K$_2$CoF$_4$ and RbCoF$_3$. However, the magnitude value of $\Gamma^i$ in Y$_2$Co$_3$ is much larger than those of K$_2$CoF$_4$ and RbCoF$_3$, where $\Gamma^i$ is around 15~\cite{Shrivastava1976}. This larger value is rationalized by the larger interlayer Co-Co distance in Y$_2$Co$_3$ (4.75\,\AA) compared to the distance between Co$^{2+}$ cations in K$_2$CoF$_4$ and RbCoF$_3$ ($\sim $\,\AA). These materials have lower $T_\textrm{N}$ values and smaller Gr\"{u}neisen parameters. Consequently, the effect of thermal contraction on their exchange interactions is less pronounced than in Y$_2$Co$_3$. We note that the temperature dependence of $\chi_\perp$ in K$_2$CoF$_4$ has been explained previously using the perturbation of the Ising model~\cite{Fisher1963Perpendicular}, but the proposed 2D Ising model is not applicable to Y$_2$Co$_3$. It is possible that the lattice contraction effects also play a role in K$_2$CoF$_4$.

The itinerant nature of Y$_2$Co$_3$ cannot be fully described by the localized superexchange model. Thus, the impact of lattice contraction on the exchange coupling in Y$_2$Co$_3$ is further compared to other Y-Co compounds with itinerant behavior. In the Y-Co system, pressure studies have systematically investigated the correlation between Curie temperature $T_\textrm{C}$ and lattice constants~\cite{Brouha1973Pressure}. Table~\ref{table} shows the pressure dependencies of $T_\textrm{C}$, the values of $\Gamma^V =- \frac{d\ln T_\textrm{C}}{d\ln V}$, and the compressibilities ($\kappa$) of various Y-Co compounds, revealing a clear trend: a lower Co content correlates with higher compressibility, reduced $T_\textrm{C}$, a greater sensitivity of $T_\textrm{C}$ to pressure $\left(\frac{dT_\textrm{C}}{dp}\right)$, and thus a larger $\Gamma^V$. Among these compounds, YCo$_3$ has the highest $\Gamma^V$ value of $-$15.1. Taking the relationship that $d\ln V= \sum_id\ln a_i$, the Gr\"{u}neisen parameter along each axis $\Gamma^i$ is approximately three times that of $\Gamma^V$, resulting in $\Gamma^i\approx -45$ for YCo$_3$. This value is comparable to that of Y$_2$Co$_3$, despite the different signs owing to the different types of exchange involved.

%\begin{table*}[!htb]
%\centering
%\caption{The transition temperature $T_\textrm{C}$ or $T_\textrm{N}$, compressibility $\kappa$, pressure dependence of transition temperature $\frac{dT_\textrm{C}}{dp}$, and the Gr\"{u}neisen parameter $\Gamma^i$ of Y-Co compounds.
%\label{table}}
%\includegraphics[width=\linewidth]{Gruneisen_table.eps}
%\end{table*}

\begin{table*}[!ht]
\caption{The transition temperature $T_\textrm{C}$ or $T_\textrm{N}$, compressibility $\kappa$, pressure dependence of transition temperature $\frac{dT_\textrm{C}}{dp}$, and the Gr\"{u}neisen parameter $\Gamma^i$ of Y-Co compounds.\label{table}}	
\renewcommand{\arraystretch}{2.0}
\begin{tabular}{>{\centering}m{2cm} >{\centering}m{1cm} >{\centering}m{2cm} >{\centering}m{2.5cm} >{\centering}m{1cm} >{\centering}m{1.5cm} >{\centering}m{1.5cm} >{\centering}m{1.5cm} >{\centering}m{1cm} >{\centering}m{1.5cm}}
\hline\hline 
Compound & Type & $T_\textrm{C}$ or $T_\textrm{N}$ (K) & $\kappa$ (10$^{-3}$ GPa$^{-1}$) & \multicolumn{3}{c}{$\dfrac{dT_\textrm{C}}{dp}$ (K GPa$^{-1}$) }& \multicolumn{2}{c}{$\Gamma^V_\textrm{FM}$} & Ref \tabularnewline  \hline
Y$_{2}$Co$_{17}$  & FM  & 1167 & 6.7 &  \multicolumn{3}{c}{$-$3}  & \multicolumn{2}{c}{$-0.4$} &\cite{Brouha1973Pressure} \tabularnewline  
YCo$_{5}$         & FM  & 977  & 7.5 &  \multicolumn{3}{c}{$-$10}  & \multicolumn{2}{c}{$-1.6$} &\cite{Brouha1973Pressure} \tabularnewline 
Y$_{2}$Co$_{7}$   & FM  & 639  &     & \multicolumn{3}{c}{$-$33}  & \multicolumn{2}{c}{$-6.4$} &\cite{Brouha1973Pressure} \tabularnewline 
YCo$_{3}$         & FM  & 301  & 8.5 & \multicolumn{3}{c}{$-$38}  &	\multicolumn{2}{c}{$-15.1$}  & \cite{Brouha1973Pressure} \tabularnewline 
YCo$_{2}$         & PM  &     & 9.8 & \multicolumn{3}{c}{ }     &	\multicolumn{2}{c}{ } & \cite{Brouha1973Pressure} \tabularnewline  \hline
\multirow{2}*{Y$_{2}$Co$_{3}$}   & \multirow{2}*{AFM} & \multirow{2}*{252}  & \multirow{2}*{12.4}    & $\dfrac{dT_\textrm{N}}{dp}$ & $\dfrac{d(C\lambda_s/2)}{dp}$ &  $\dfrac{d(C\lambda_d/2)
}{dp}$ &$\Gamma_\textrm{FM}^{ac}$   & $\Gamma_\textrm{AFM}^b$ & \multirow{2}*{This paper} \tabularnewline  \cline{5-9}

                  &     &     &       &   $-$1.65   &   $-$24.65   &  $-$23 &$-$23 & 66 &    \tabularnewline  \hline\hline
\end{tabular}
\end{table*}

Notably, the pressure dependence of $T_\textrm{N}$ $\left(\frac{dT_\textrm{N}}{dp} = -1.65 \, \text{K/GPa}\right)$~\cite{ShiRobust2021PRB} in Y$_2$Co$_3$ is markedly lower compared to $\frac{dT_\textrm{C}}{dp}$ in other Y-Co compounds, despite similar compressibilities observed in our pressure studies of Y$_2$Co$_3$ (see Appendix~A). This is because $T_\textrm{N} =  \frac{C}{2}(\lambda_s-\lambda_d)$ is determined by both the intraplane FM and interplane AFM interactions. Lattice contraction under pressure suppresses ferromagnetic exchange within the kagome plane, which is consistent across other Y-Co compounds while enhancing antiferromagnetic exchange caused by reduced interplane separations. Thus, the pressure dependence of these two components needs to be analyzed separately. The influence of pressure on the AFM exchange in Y$_2$Co$_3$ is quantitatively assessed through $\Gamma^b$ using the equation $\frac{1}{C\lambda_d}\frac{d(C\lambda_d)}{dp} = \kappa_b\Gamma_{\text{AFM}}^b$, where $\kappa_b$ denotes the compressibility along the $b$ axis. This yields $\frac{d(C\lambda_d/2)}{dp} = -23$\,K/GPa. The FM interaction dependence on pressure can then be calculated by $\frac{d(C\lambda_s/2)}{dp} = \frac{dT_\textrm{N}}{dp}+\frac{d(C\lambda_d)/2}{dp}=-24.65$ \,K/GPa, comparable to that of other Y-Co compounds with itinerant ferromagnetic ordering.  Consequently, $T_\textrm{N}$ exhibits minimal variance under pressure as a result of compensating FM and AFM interactions. The Gr\"{u}neisen parameter of the FM exchange interaction within the $ac$ plane, denoted as $\Gamma_{\textrm{FM}}^{ac}$, can be similarly calculated using the equation $\frac{1}{C\lambda_s}\frac{d(C\lambda_s)}{dp} = \kappa_{ac}\Gamma_{\text{FM}}^{ac}$, yielding a value of approximately $-$23.

\subsection{The evidence for itinerant  behavior from physical property measurements}

The decrease of the perpendicular magnetic susceptibility is well explained by the increase of the AFM exchange interaction at lower temperatures. However, the origin of the nonzero parallel magnetic susceptibility remains an open question. The magnetic structure of this compound is complicated because of the existence of competing ferromagnetic and antiferromagnetic interactions of the $3d$ electrons in the regime between local and itinerant characters. It has been seen in TiAu, an itinerant antiferromagnetic compound without magnetic constituents, that the magnetic susceptibility shows typical AFM behavior with a value of $10^{-3}$\,e.m.u./mol at the low temperature while displaying the Curie-Weiss behavior above the transition temperature~\cite{Svanidze2015}. The residual magnetic susceptibility at low temperature of TiAu is comparable to that of Y$_2$Co$_3$. In this paper, we propose that the itinerant AFM component may contribute to the nonzero magnetic susceptibility seen in Y$_2$Co$_3$. Here, we show some evidences of the itinerant AFM component from both Hall measurements and heat capacity results.

\begin{figure}[!htb]
	\centering
	\includegraphics[width=\linewidth]{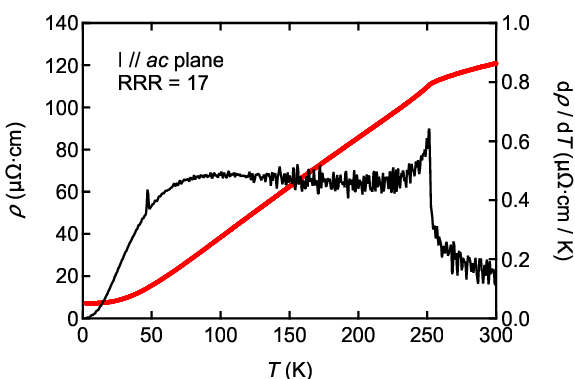}
	\caption{Resistivity of Y$_2$Co$_3$ and its derivative as a function of temperature. A sudden change of slope corresponding to the AFM transition at 252\,K is observed.
	\label{Resistivity_ac.eps}} 
\end{figure}

Figure~\ref{Resistivity_ac.eps} displays the transport properties of Y$_2$Co$_3$. The comparatively low resistivity signifies good metallic behavior. The derivative (black curve) exhibits a sudden jump at $T_\textrm{N}$, corresponding to the reduction of scattering from the disordered state to the ordered state. Notably, this compound does not exhibit the typical resistivity increase owing to gap opening associated with spin density waves. However, similar decreases in resistivity have also been observed in other itinerant antiferromagnetic (IAFM) compounds containing $d$ electrons, such as CaCo$_2$P$_2$~\cite{Suppression2014Baumbach}, TiAu~\cite{Svanidze2015}, and BaFe$_2$As$_2$~\cite{Spin2008Rotter}.

% Figure\ref{Resistivity_ac.eps} exhibits the transport properties of Y$_2$Co$_3$. The relatively low resistivity indicates good metal behavior. The derivative (black curve) shows a sudden jump at $T_\textrm{N}$, which is corresponding to the decrease of scattering from the disordered state to the ordered state. In this compound we don't see the typical increase of resistivity due to the gap opening associated with the spin density wave, however, a similar drop of resistivity is also observed in other IAFM compounds with d electrons, such as CaCo$_2$P$_2$~\cite{Suppression2014Baumbach}, TiAu~\cite{Svanidze2015} and BaFe$_2$As$_2$~\cite{Spin2008Rotter}.

\begin{figure}[!htb]
	\centering
	\includegraphics[width=\linewidth]{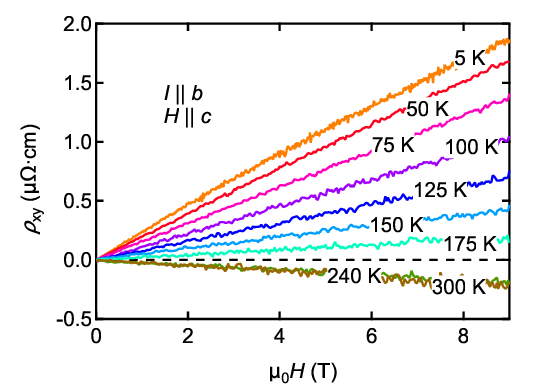}
	\caption{The Hall resistivity as a function of applied magnetic field along the $c$ axis at various temperatures.
	\label{Hall_Res}} 
\end{figure}

Figure~\ref{Hall_Res} shows the magnetic field dependence of the Hall resistivity at various temperatures from 5\,K to 300\,K. The Hall coefficient shows a strong temperature dependence and undergoes a change in sign at approximately 200\,K. The charge density is estimated to be $n_h=2.75\times10^{21}$ cm$^{-3}$ for holes at 5\,K and $n_e=2.52\times10^{22}$ cm$^{-3}$ for electrons at 300\,K. The temperature-dependent sign reversal of the Hall coefficient clearly indicates multiband contributions and change of the band structure. It is worth noting that this sign reversal occurs in close proximity to $T_\textrm{N}$, suggesting that the change in the band structure near the Fermi level is likely induced by the magnetic ordering. This observation provides further support for the non-negligible influence of an itinerant antiferromagnetic component.

\begin{figure}[!htb]
	\centering
	\includegraphics[width=\linewidth]{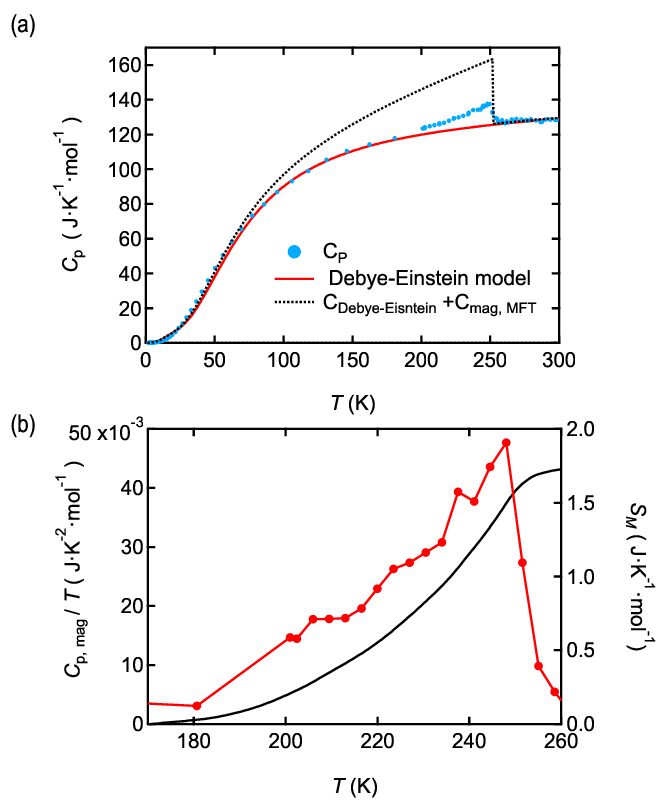}
	\caption{(a) Specific heat of Y$_2$Co$_3$. The solid-red line is the calculated lattice specific heat using the Debye-Einstein model. The dashed-black line is the total calculated specific heat. The magnetic specific heat is calculated following the MFT model. (b) Magnetic specific heat and magnetic entropy $S_M$.
    	\label{HC_Sm}}
\end{figure}

The presence of the itinerant AFM component is also supported by the small magnetic entropy $S_M$ near the transition. Figure~\ref{HC_Sm}(a) shows the specific heat of Y$_2$Co$_3$ as a function of temperature. The lattice specific heat is estimated with the Debye-Einstein model with three acoustic branches and $(3n-3)$ optical branches, where $n=10$ is the number of atoms in one primitive cell. The Debye temperature and Einstein temperature are $T_\Theta=91$\,K and $T_E=225$\,K, respectively. A fit with Debye only gave unphysical results reflecting the importance of the optical phonon modes. The specific heat of the magnetic ordering is calculated following the MFT model~\cite{JohnstonPRB2011Magnetic} and compared with the experiment data. By subtracting the lattice specific heat, one can obtain the magnetic specific heat across the transition. The magnetic entropy change is estimated to be 1.7\,J/(K mol) or 0.56\,J/K per Co, which is only 10\% of $R\ln 2$.

\section{Spin-Polarized DFT calculations}

In order to shed light on the magnetic behavior described in the above experiments we have conducted detailed calculations of the magnetic ground state (moments parallel to the $b$ axis, successive kagome planes anti-aligned) depicted in Fig.~\ref{MagneticStructure_Neutron}(a). Details of these calculations may be found in our previous publication~\cite{ShiRobust2021PRB}.

In general, the calculations do a rather good job of describing the observed magnetic ground state, at least with respect to the collinear component considered here. We find moments on the $8e$ and $4b$ sites of, respectively, 0.84 and 1.43\,$\mu_B$/Co, in reasonable agreement with the experimental values of 0.68(3) and 1.25(4)\,$\mu_B$/Co. The larger theoretical values may originate in any of a number of factors. These would include the standard first-principles neglect of fluctuations, and our use of the generalized gradient approximation, which for itinerant systems can overstate magnetic order relative to the local density approximation. Nevertheless, we reproduce both the relative moment size ordering of the two distinct Co sites in this complex structure, the general magnitude of these moments, and the energetic favorability of the ground state itself, relative to other states such as a ferromagnetic configuration, which falls some 6\,meV/f.u higher in energy.

In addition, our calculations find the $b$ axis as the easy\,axis, again as found experimentally, with this axis more stable than the $a$ axis by an energy value of 0.15\,MJ/m$^{3}$ and the corresponding value with respect to the $c$ axis rather larger at 0.90\,MJ/m$^3$. This is contrary to the experimental result where we find a small canting along $c$, but not along $a$. This highlights the limits of DFT calculations for such small energy scales. These two values bracket the experimental value for hcp, {\it ferro}magnetic cobalt of approximately 0.7\,MJ/m$^{3}$~\cite{paige}. 
 These values may be regarded as relatively typical for 3$d$-based compounds such as Y$_{2}$Co$_{3}$, and rather smaller than those, such as YCo$_5$, of the CaCu$_{5}$ structural type, which display an order of magnitude larger magnetic anisotropy.

 These typical anisotropy values in Y$_2$Co$_3$ arise despite the presence here of kagome-type planes, although highly distorted, as in the CaCu$_5$ structural type and speak to the importance of both the intervening (between Co layers) yttrium plane, and the particular stoichiometry of Y$_2$Co$_3$ in determining the magnetic ground state and its interaction with the lattice.

 We also note from Fig.~\ref{MvH_60T_all}(b) (see Appendix~D) that the experimentally observed exchange field -- that required to fully polarize the material into a ferromagnetic state -- is likely well in excess of 100\,T. This indicates that the magnetism in this compound is in fact rather robust, as found in our previous publication. We note that this occurs despite the fact that both the calculated and observed magnetic moment values are significantly smaller than those in cobalt itself, indicative of the complexity of the magnetic interactions in Y$_{2}$Co$_{3}$.

 Finally, note that, as detailed in our previous publication on Y$_2$Co$_3$, there is significant evidence even from the first-principles calculations of the itinerant character described experimentally above. In particular, certain magnetic configurations,  when initialized (such as a {\it ferrimagnetic} state described in that work), spontaneously change in the course of the calculation convergence to a different magnetic configuration, while others -- such as one with Co-Co nearest-neighbors anti-aligned -- simply fail to converge. Both these results are strong theoretical indications that the magnetism, while clearly containing elements of local character, also possesses a degree of itinerant behavior. This is indeed a complex behavior that could not have reasonably been anticipated, and suggests that even among materials one would expect to be easy to describe, unanticipated surprises may be present.

\section{Conclusion}
In summary, this study presents the magnetic structure of Y$_2$Co$_3$. single-crystal neutron diffraction reveals that Y$_2$Co$_3$ forms an almost collinear A-type antiferromagnetic structure, with ferromagnetic interaction within the distorted kagome plane and antiferromagnetic interaction between planes. The magnetic moments on the two distinct Co sites are $[0, -0.68(3), 0]$\,$\mu_B$ and $[(0, 1.25(4), 0.07(1)]$\,$\mu_B$, respectively. We addressed the discrepancy between the almost collinear magnetic structure and the anomalous temperature dependence of the bulk magnetization within the local and itinerant regimes. We show that the AFM exchange interactions are strongly affected by lattice constants, resulting in a large temperature dependence of the perpendicular magnetic susceptibility. The itinerant AFM component may explain the nonzero parallel magnetic susceptibility at low temperatures,  as indicated by resistivity measurements and the minimal entropy change observed in heat capacity measurements. First-principles calculations support these results.

\section*{Acknowledgement}
We thank D. C. Johnston for useful discussions. Y.S. and V.T. acknowledge support from the UC Laboratory Fees Research Program (LFR-20-653926) and the UC Davis Physics Liquid Helium Laboratory Fund. Part of this research used resources at the High Flux Isotope Reactor, a DOE Office of Science User Facility operated by the Oak Ridge National Laboratory. Work done at Ames National Laboratory (SLB and PCC) was supported by the U.S. Department of Energy, Office of Basic Energy Science, Division of Materials Sciences and Engineering.  Ames National Laboratory is operated for the U.S. Department of Energy by Iowa State University under Contract No. DE-AC02-07CH11358. The experimental research at the University of Utah received support from the National Science Foundation Division of Materials Research Award No. 2132692. This work was also supported by the U.S. Department of Energy (DOE) Office of Science, Fusion Energy Sciences funding award entitled High Energy Density Quantum Matter, Award No DE-SC0020340. The experimental work took place at HPCAT (Sector 16, Advanced Photon Source, Argonne National Laboratory), which is supported by the DOE/National Nuclear Security Administration, Office of Experimental Sciences, with x-ray beam time and personnel support made possible by the Chicago/DOE Alliance Center, which is funded by the DOE/NNSA (DE-NA0003975). Instrumentation and facilities used were also supported by the National Science Foundation (DMR-1809783). The Advanced Photon Source is operated by the DOE Office of Science by Argonne National Laboratory under Contract No. DE-AC02-06CH11357. The work of DP and LY at Oak Ridge National Laboratory (first principles calculations and interpretation) was supported by the U.S. Department of Energy, Office of Science, Basic Energy Sciences, Materials Science and Engineering Division. Measurements in pulsed magnetic fields were funded by DOE Basic Energy Sciences program ``Science of 100 tesla."

\section*{Appendix A: The Crystal Structure of Y$_2$Co$_3$ under high pressure} 
We  investigated the pressure effects on the structural properties of Y$_2$Co$_3$ up to 19\,GPa. High-pressure diffraction data were collected at the 16-BM-D beamline of the High Pressure Collaborative Access Team (HPCAT) at APS, ANL ($\lambda$ = 0.4133 Å). Boehler-Almax plate diamond-anvil cells (DACs) with 350 or 500 \textmu m culet size and a large opening angle of approximately 60° were utilized. The experiments involved high-pressure powder x-ray measurements conducted in two separate runs. All measurements were done at room temperature. Data collection between 3.65$-$9.8\,GPa utilized silicon oil as the pressure-transmitting medium (PTM), while the second run, between 7.3$-$21\,GPa, employed isopropanol as the PTM. We loaded the sample, along with ruby and gold, into a stainless steel gasket, which was then filled with the PTM. Pressure was determined using the ruby fluorescence method and the equation of state (EOS) of Au~\cite{Anderson1989JAP}. The diffraction images were integrated using the Dioptas program~\cite{Prescher2015DIOPTAS}. The powder XRD data were analyzed by the Le Bail fitting method using GSAS-EXPGUI package~\cite{Lake_Toby_2011}. 

\begin{figure}[!htb]
	\centering
	\includegraphics[width=\linewidth]{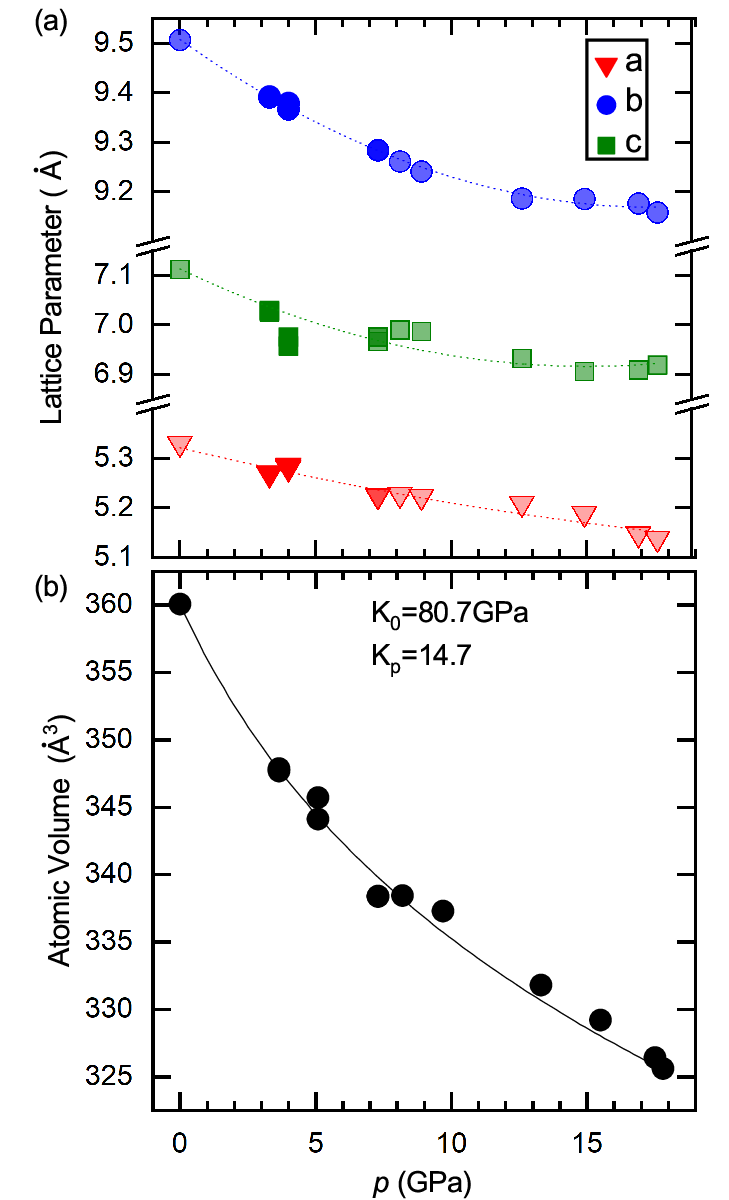}
	\caption{(a) The lattice parameters $a$, $b$, $c$ as a function of applied pressure. The dashed lines are polynomial fittings. (b) The unit-cell volume as a function of pressure. The solid line shows fourth-order Birch–Murnaghan fit to the experimental data.
	\label{lattice_parameter_pressure}}
\end{figure}

\begin{figure}[!htb]
	\centering
	\includegraphics[width=\linewidth]{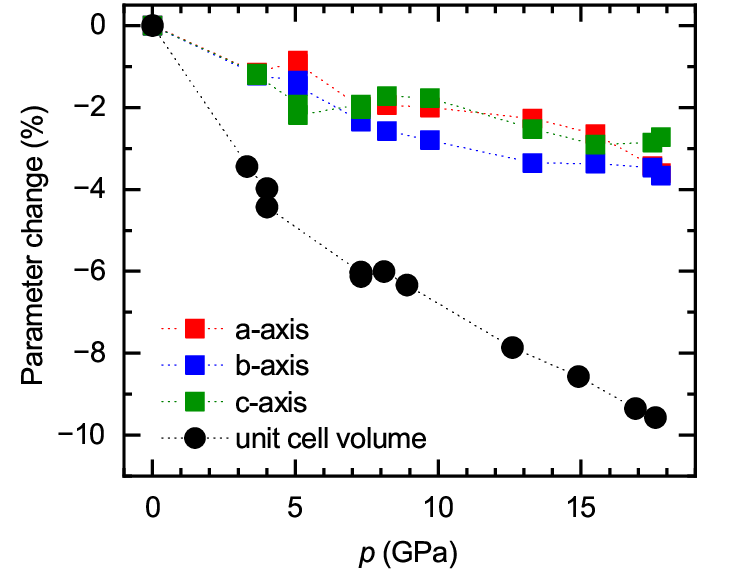}
	\caption{Percentage change in lattice constants and unit-cell volume as a function of pressure.\label{percentage change}}
\end{figure}

No structural phase transition was observed up to 21\,GPa, and the sample remained in its ambient pressure $Cmce$ phase~\cite{ShiRobust2021PRB}. The sample exhibited a large bulk modulus of 80.7 GPa, which was determined with EoSFit7\_GUI (Fig.~\ref{lattice_parameter_pressure})~\cite{ShenToward2020}. The percentage of axial compression was similar along all axes but might be slightly higher along the $b$ axis (Fig.~\ref{percentage change}). The $b$ axis separates layers of Co and Y and controls the corresponding interlayer AFM coupling, which is expected to be strengthened under compression and help enhance the N\'{e}el temperature. Meanwhile, all primary intralayer FM couplings are weakened because of the compression along the $a-c$ planes. The overall changes in the magnetism of Y$_2$Co$_3$, therefore, represent a balance of these two effects. Low-pressure data show that below 1\,GPa, the N\'{e}el temperature decreases by a small amount of 1.6\,K which is consistent with the similar compression rates of the sample along different axes.

%Figure~\ref{lattice_parameter_pressure} shows the pressure dependence of the lattice parameters and the unit cell volume. In the low-pressure regime, $dV/dp=-3.08$\,\AA$^3$/GPa, giving the value of $\kappa=8.6\times10^{-4}$\,kbar$^{-1}$. As the pressure becomes larger, the slope of the curve becomes smaller. The compressibility of Y$_2$Co$_3$ listed in table~\ref{table} is obtained from the low-pressure region in order to compare with other Y-Co compounds under similar pressures.

\section*{Appendix B: The Sublattice Magnetic Moment and Perpendicular Magnetic Susceptibility}

Following the molecular field theory, the sublattice magnetization at zero field can be calculated by solving the Brillouin function~\cite{Blundell2001Magnetism1, Johnston2012MagneticPRL, JohnstonPRB2011Magnetic}

\begin{subequations}
 \begin{eqnarray}
  M_1 = M_{sat}B_s\left(\frac{g\mu_BS}{k_BT}(\lambda_sM_1+\lambda_dM_2)\right) \label{eqn: 5a} \\
  M_2 = M_{sat}B_s\left(\frac{g\mu_BS}{k_BT}(\lambda_sM_2+\lambda_dM_1)\right)\label{eqn: 5b}
 \end{eqnarray}
\end{subequations}

In these equations, $M_1$ and $M_2$ are the magnetizations of two sublattices with opposite signs. $M_{sat} = \frac{1}{2}Ng\mu_BS$ is the saturated magnetization of one subllattice. Under zero field, the magnitudes of the magnetization of two sublattices are the same. Thus, Eqs.~(\ref{eqn: 5a}) and (\ref{eqn: 5b}) can be simplified as
\begin{equation}
    \frac{M}{M_{sat}} = B_s\left(\frac{g\mu_BS}{k_BT}(\lambda_s-\lambda_d)M\right) %=B_s\left(\frac{3S}{(S+1)t}\frac{M}{M_{sat}}\right)
\end{equation}

The temperature dependence of reduced sublattice magnetic moment can then be calculated with and without $\lambda_d$ changing with temperature.  In the calculation, $\lambda_s$ is a constant because the area of the $ac$ plane almost remains temperature independent.  

\begin{figure}[!htb]
 	\centering
 	\includegraphics[width=\linewidth]{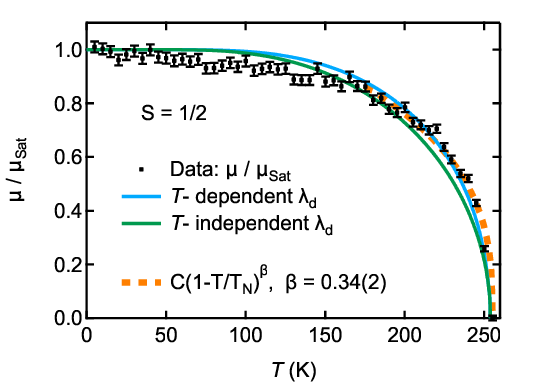}
 	\caption{The normalized ordered moment $\mu(T)/\mu_{sat}$ compared with the MFT calculations (solid lines) for spin $S = 1/2$. The orange-dash line is the fitting near the transition using the power law. \label{ordered_moment_MFT}}  
\end{figure}

The comparison between the MFT calculation and the temperature-dependent ordered moment from neutron diffraction is shown in Fig.~\ref{ordered_moment_MFT}. The ordered moment is obtained from the square root of the Bragg peak intensity after the removal of the background signal above the transition temperature~\cite{EuMg2Sb2Pakhira}. Near the transition temperature, the MFT prediction with the temperature-dependent molecular field $\lambda_d$ fits the experiment better. At low temperatures, both calculation curves can not describe the nonsaturation behavior, which is likely caused by the magnon excitation that is not discussed in the MFT model. A power law $C(1-\frac{T}{T_\textrm{N}})^\beta$ is used to fit the ordered moment near the transition, which gives a $\beta$ value of 0.34(2), relatively close to the critical exponent of the Heisenberg model, which is 0.367~\cite{Blundell2001Magnetism1}.

As the magnetic field is applied perpendicular to the direction of the magnetic moments, the magnetization of each sublattice tilts slightly towards the direction of the magnetic field, producing a nonzero magnetic susceptibility. The tilted angle $\theta$ is determined by the competition between the AFM exchange energy, the Zeeman energy, and the magnetocrystalline anisotropy energy. The angle between the sublattice magnetization and the magnetic field is written as $\phi=90\degree-\theta$, and the angle between the two sublattices magnetization is $2\phi$.

The total energy of the two sublattices in a perpendicular magnetic field can be written as

\begin{equation}
    E=-2MB\cos\phi+|\lambda_d|M^2\cos2\phi-K(1-\cos^2\phi)
\end{equation}
In this equation, $M$ is the magnetization of each sublattice.
Taking $\frac{\partial E}{\partial \phi}=0$, the energy is minimized when $\cos\phi=\frac{MB}{2|\lambda_d|M^2+K}$. Thus, the total perpendicular magnetization is 
\begin{equation}
    M_{\perp}=2M\cos\phi=\frac{2M^2B}{2|\lambda|_dM^2+K}
\end{equation}
and the perpendicular magnetic susceptibility is 
\begin{equation}
    \chi_\perp=\frac{2M^2}{2|\lambda_d|M^2+K} \label{eqn: chi_perp}
\end{equation}

As the magnetocrystalline anisotropy energy is much smaller than the exchange energy, Eq.~(\ref{eqn: chi_perp}) can be simplified as $\chi_\perp=\frac{1}{|\lambda_d|}$.
\section*{Appendix C: Estimation of Canting angle in the Non-collinear Model}
\begin{figure}
    \centering
    \includegraphics[width=\linewidth]{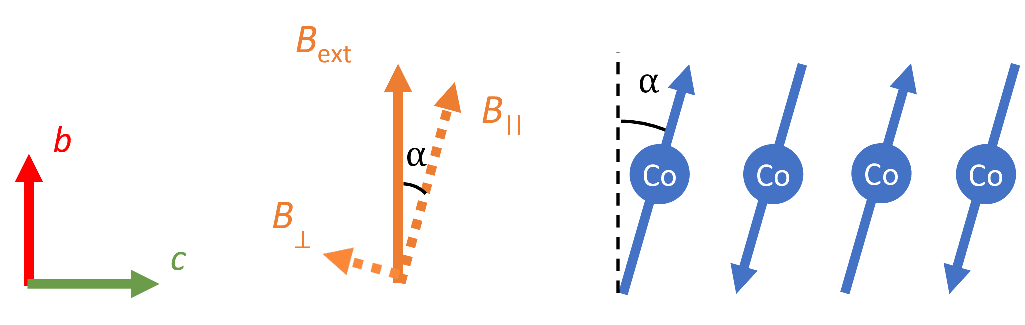}
    \caption{An illustration of a simple canting model. The spins are anti-aligned and canted from the $b$ axis toward the $c$ axis by an angle of $\alpha$. The external magnetic field is decomposed into a parallel and a perpendicular components relative to the direction of the magnetic moment.}
    \label{fig:canting_angle_estimation}
\end{figure}

We derive the canting angle based on a simple canting model shown in Fig.~\ref{fig:canting_angle_estimation}: a pair of spins are anti-aligned and canted from the $b$ axis towards the $c$ axis by an angle of $\alpha$. 

When an external magnetic field $B$ is applied perpendicular to the magnetic moment, the susceptibility below $T_\textrm{N}$ is independent of the temperature, $\chi = \chi(T_\textrm{N})$. When the magnetic field is applied parallel to the direction of the magnetic moment, the magnetic susceptibility decreases from $\chi(T_\textrm{N})$ to 0 as the temperature decreases~\cite{Blundell2001Magnetism1}. 

For a sample with a magnetic field applied along the $b$ axis, the applied field has both parallel and perpendicular components relative to the moment direction. At low temperatures, the parallel component of the magnetic field yields zero magnetization, and the perpendicular magnetic field $B_{\perp} = B\sin{\alpha}$ would produce a magnetization
\begin{equation}
    M(T=0) = \chi(T_\textrm{N})B_{\perp} = \chi(T_\textrm{N})B\sin\alpha
\end{equation}

Since our instrument measures the magnetization along the direction of the magnetic field, the measured $b$ axis magnetization is
\begin{equation}
    M_{b}(T=0)=M(T=0)\sin\alpha = \chi(T_\textrm{N})B\sin^2\alpha
\end{equation}

Therefore, the measured susceptibility at zero temperature is
\begin{equation}
    \chi_{b}(T=0) = \frac{M_{b}(T=0)}{B} = \chi(T_\textrm{N})\sin^2\alpha
\end{equation}

This yields $\sin^2\alpha = \dfrac{\chi_{b}(T=0)}{\chi(T_\textrm{N})}$.

\section*{Appendix D: The Spin-Flop Phase Diagram and Temperature Dependence of Magnetocrystalline Anisotropy}
\begin{figure}[!htb]
	\centering
	\includegraphics[width=\linewidth]{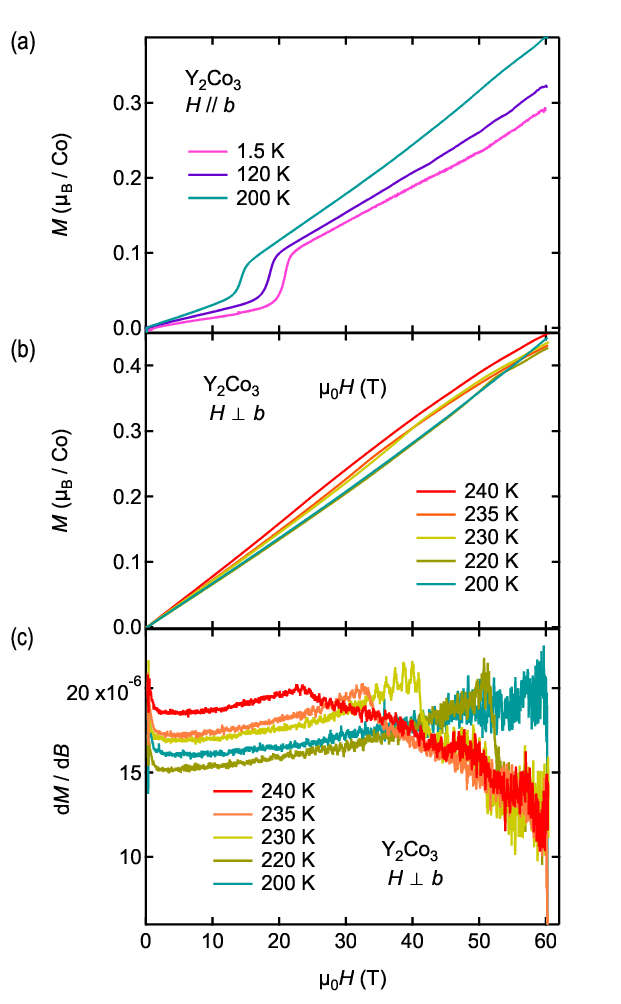}
 	\caption{(a) Magnetization as a function of applied magnetic field along the $b$ axis and (b) in the $ac$ plane. (c) Derivative of magnetization vs magnetic field.
 	\label{MvH_60T_all}}  
\end{figure}

% \begin{figure}[!htb]
% 	\centering
% 	\includegraphics[width=\linewidth]{Y2Co3_phasediagram_para.eps}
% 	\caption{An illustration of the magnetic phase diagram with the magnetic field along the $b$ axis. The red solid circle is from the DC pulse field result. The red solid square is from the AC susceptibility result~\cite{ShiRobust2021PRB}. The blue solid circle is from the resistivity measurement using PPMS.
% 	\label{phase_diagram}}  
% \end{figure}

\begin{figure}[!htb]
	\centering
	\includegraphics[width=\linewidth]{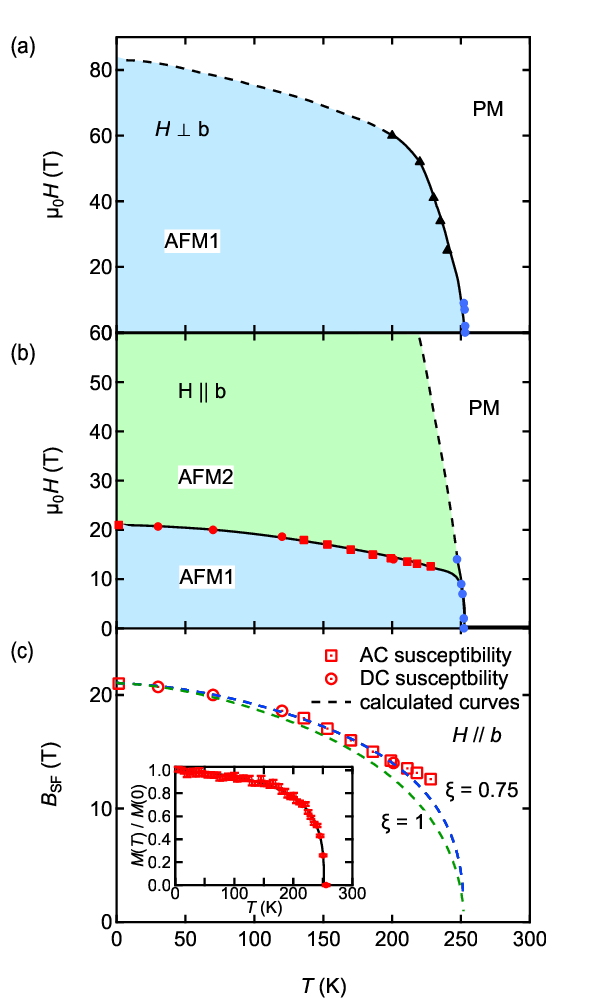}
	\caption{(a) An illustration of the magnetic phase diagram with the magnetic field perpendicular to the $b$ axis. The blacktriangle is from the DC pulse field result. The blue-solid circle is from the resistivity measurement using PPMS. (b) The magnetic phase diagram with the magnetic field along the $b$ axis. The blue region represents the original antiferromagnetic phase (AFM1), while the green region shows the spin-flop phase (AFM2). The red-solid circle is from the DC pulse field result. The red-solid square is from the AC susceptibility result~\cite{ShiRobust2021PRB}. The blue-solid circle is from the resistivity measurement using PPMS. (c) The spin-flop phase diagram with $B_{SF}$ as a function of temperature. The dashed line is the calculated curve of the phase boundary. The inset figure is the normalized magnetization of one sublattice below transition, obtained from the neutron magnetic Bragg peak intensity. The solid line is the fitting curve using the Kuz'min equation~\cite{Shape2005Kuzmin}.
	\label{Bsf_simulation}} 
\end{figure}

Y$_2$Co$_3$ undergoes a spin-flop transition when the magnetic field is applied along the $b$ axis at 21\,T and 2\,K~\cite{ShiRobust2021PRB}. In this study, we have taken a more comprehensive examination of the spin-flop transition phase diagram throughout the entire temperature range below $T_\textrm{N}$, as is shown in Fig.~\ref{Bsf_simulation}. We show a relatively good alignment between the experiment data and the theoretical calculation of the phase boundary based on a local moment collinear AFM model. 

The high magnetic field measurements were performed using an extraction magnetometer in pulsed fields, with the sample immersed either in helium gas ($T>4$\,K) or $^4$He  liquid ($T<4$\,K). The extraction magnetometer enables measurements to be performed with and without the sample present to enable subtraction of the background~\cite{Detwiler2000Magnetization}. 

The field dependence of magnetization $M(H)$ up to 60\,T is shown in Fig.~\ref{MvH_60T_all}, without reaching saturation in both directions. Despite the relatively large effective moment of 2.5 $\mu_B$/Co derived from the Curie-Weiss fitting in the high-temperature paramagnetic regime, the field-dependent magnetization remains small, indicating that the AFM ordering is very robust.  When the magnetic field is applied perpendicular to the easy axis [see Fig.~\ref{MvH_60T_all}(b)], an antiferromagnetic to paramagnetic transition occurs, which is more apparent in the derivative $dM/dB$ curves in Fig.~\ref{MvH_60T_all}(c). 

When the magnetic field is applied along the $b$ axis, a spin-flop transition is observed at various temperatures [see Fig.~\ref{MvH_60T_all}(a)], in good agreement with previously reported results~\cite{ShiRobust2021PRB}. Such spin-flop transition corresponds to the reorientation of local moments from the $b$ axis to the $ac$ plane because of the competition between the Zeeman energy, exchange energy, and the magnetocrystalline anisotropy energy. 
In this paper, we simplify the magnetic structure to an A-type AFM structure and disregard the distinctions between the two cobalt sites. The temperature-dependent spin-flop field can be determined by the following equation~\cite{ShiRobust2021PRB}:

\begin{equation}
    B_{SF}(T)=\sqrt{\frac{K(T)(2|\lambda_d|M^2-K(T))}{M^2}}
\end{equation}

Here, $K(T)$ is the temperature-dependent anisotropy energy, $|\lambda_d|M^2$ represents the exchange energy between two sublattices, and $M$ is the sublattice magnetization per Co. We used $M_{sat}(T=0)=1\,\mu_B$\ ($S=\frac{1}{s}$) in our simulation. At 2\,K, $B_{SF}(0)$ and $K(0)$ were previously reported to be 21\,T and $11.46$\,J per mol of pair of Co sites (or 5.73\,J/mol-Co), respectively. The exchange energy term $|\lambda_d|M^2$ can also be determined to be 303\,J per mol of Co. Converting this to $C\lambda_d$ at 2\,K, we get $-$304\,K, further confirming that the exchange energy is enhanced as the temperature decreases.

The temperature dependence of $K(T)$ is given by the Callen-and-Callen (CC) law~\cite{CALLEN1960Anisotropic, Callen2004TEMPERATURE, Lamichhane2020Reinvestigation}. In a uniaxial system, the temperature-dependent anisotropy can be written as

\begin{equation}
K_1(T)=(K_1^0+\frac{7}{8}K_2^0)\left(\frac{M(T)}{M(0)}\right)^3-\frac{7}{8}K_2^0\left(\frac{M(T)}{M(0)}\right)^{10}\\
\end{equation}

\begin{equation}
K_2(T)=K_2^0\left(\frac{M(T)}{M(0)}\right)^{10}
\end{equation}

Considering that the $K_2$ term is usually negligible, the equation can be reduced to
\begin{equation}
    K(T)=K_0\left(\frac{M(T)}{M(0)}\right)^3  \label{K_power_law}
\end{equation}

In an antiferromagnet, however, it is not rigorous to directly apply this power law by taking $M(T)$ as the sublattice magnetization. Equation~(\ref{K_power_law}) is modified as~\cite{Callen1966Present}
\begin{equation}
    K(T)=K_0\left(\frac{M(T)}{M(0)}\right)^{3\xi}  \label{K_power_law_modif}
\end{equation}
Note that the equation above is obtained based on a particular model with a single-ion mechanism spin-wave calculation, and is not accurate in a complicated magnetic system such as Y$_2$Co$_3$. Thus, this is only a rough representation of the experimental data.
In this paper, the $\frac{M(T)}{M(0)}$ of Y$_2$Co$_3$ is determined by the Bragg peak intensity of single-crystal neutron scattering, as mentioned in Appendix B. The curve is fitted using the so-called Kuz'min equation, as shown in Fig.~\ref{Bsf_simulation}(a),

\begin{equation}
    m(\tau)=[1-s\tau^{3/2}-(1-s)\tau^p]^{1/3}
\end{equation}

In this equation, $m=M_s/M_0$ is the normalized spontaneous magnetization, $\tau=T/T_\textrm{C}$ (or $T/T_\textrm{N}$ in AFM). $s$ and $p$ are the parameters that determine the shape of the magnetization curve, and are fitted to be 0.63 and 5.4, respectively. The temperature dependence of anisotropy $K(T)$ and spin-flop field $B_{SF}$ can then be represented as

\begin{equation}
    K(T) = K_0[1-s\tau^{3/2}-(1-s)\tau^p]^\xi
\end{equation}
and
\begin{equation}
\begin{aligned}
     B_{SF}(T)={} & \big\{2|\lambda_d|K_0[1-s\tau^{3/2}-(1-s)\tau^p]^\xi\\
     & -\frac{K_0^2}{M_s^2}[1-s\tau^{3/2}-(1-s)\tau^p]^{4/3\xi}\big\}^{1/2}
\end{aligned}
\end{equation}

The calculated spin-flop field as a function of temperature is shown in Fig.~\ref{Bsf_simulation}(b) and compared with the experiment data of AC susceptibility measurement (circle) and DC pulse measurement (square). The calculated phase boundary with $\xi=0.75$ is in better agreement with the experiment result below 200\,K compared to $\xi=1$. Above 200\,K the spin-flop transition becomes very broad owing to the thermal fluctuation near the transition temperature, which may cause the deviation between the experiment data and the calculated curve.

\bibliography{Y2Co3_neutron,biblio}

\end{document}